\documentstyle[graphicx,multicol,prl,aps]{revtex}

\begin{document}
\pagestyle{empty}
\begin{multicols}{2}
\narrowtext
\parskip=0cm
\noindent
{\bf Palassini and Young reply:}
In recent work\cite{PY}, we studied numerically the ground state
structure of the nearest-neighbor Ising spin glass with Gaussian couplings in
dimensions, $d=3,4$. We looked for large-scale, low-energy excitations about
the ground state in finite systems of size $N=L^d$ with periodic boundary
conditions, by applying a bulk term in the Hamiltonian which couples to the
ground state.  The results suggest that there are excitations which flip a
finite fraction of spins, have energy which varies as $L^{\theta^\prime}$ with
$\theta^\prime \simeq 0$, and have a surface with fractal dimension $d_s$ less
than $d$. The exponent $\theta^\prime$ seems to be different from the exponent
$\theta$ which determines the energy difference when the boundary conditions
are changed, e.g. from periodic to anti-periodic, since $\theta > 0$ in both
$d=3,4$.  Similar results were obtained by Krzakala and Martin\cite{km} using
a different technique. 

In their comment on Ref.~\cite{PY}, Newman and Stein\cite{ns} (referred to as
NS) have clarified the nature of these excitations 
(assuming they persist in the thermodynamic limit). They
consider the problem from the point of view of ``thermodynamic states'',
defined in a precise way following earlier work by Fisher and Huse\cite{fh}
and
themselves\cite{ns2}.
In this approach one considers a system of size $L_B$ to be
embedded in a much larger system of size $L$.
Roughly speaking, different thermodynamic
states correspond to different sets of correlation functions in the size-$L_B$
region as $L \to \infty$. ``Extremal" thermodynamic
states are called ``pure" states.
NS argue that system-size excitations with $\theta^\prime =
0$ and $d_s < d$ would not give rise to new pure states because such
excitations would ``deflect to infinity'' rather than run through the
size-$L_B$
system. Only if $d_s = d$ would system-size excitations with $\theta^\prime =
0$ give rise to new thermodynamic states. 

The conclusions of NS are consistent with our results, since we did not claim
that these excitations would give new pure states. Such excitations could
however, as NS also note, be relevant for the excitation spectrum of spin
glasses. They could also give rise to a non-trivial order parameter function
$P(q)$ on the scale of the whole system, as has been seen in Monte Carlo
simulations down to very low temperatures\cite{kpy}.

Of course, another important issue is whether our results, which emerge from
studies of small systems, continue to be valid for large sizes. 
Based on studies of very large systems in two dimensions, where however $T_c =
0$, Middleton\cite{aam} has recently suggested
that finite-size corrections may be
large in our data because we average over fluctuations of different length
scales. Middleton also proposes that our results may go over to those of the
``droplet picture''\cite{fh} in which $\theta^\prime = \theta\ ( > 0)$ (and
$d_s < d$). Our results have also been criticized in the opposite sense by
Marinari and Parisi\cite{mp} who agree with $\theta^\prime = 0\ ( \ne \theta)$,
but argue, from a different analysis and using their own data, that $d_s = d$,
which gives the ``replica symmetry breaking" picture. 

To be sure of the correct behavior in the thermodynamic limit may require the
study of sizes much larger than the sizes currently accessible to 
simulations. However, an important question
which can potentially be addressed now is why there are two different
(possibly effective) exponents, $\theta^\prime$ and $\theta$, characterizing
large-scale excitations of different geometry for the {\em same}\/ range of
sizes.

We thank D. Stein and C. Newman for a useful correspondence.

\vskip0.4cm
\noindent 
Matteo Palassini$^1$ and A.P. Young$^2$, \\
\indent {\small $^1$University of California San Francisco}\\
\indent {\small $^2$University of California Santa Cruz}

\end{multicols}
\end{document}